\documentclass[a4paper]{jpconf}
\usepackage{graphicx,amsmath,amssymb,bm,mathrsfs}

\allowdisplaybreaks 

\usepackage{color}
\definecolor{Green}{rgb}{0,0.7,0}

\newcommand{\suzum}[1]{\textcolor{black}{#1}}
\newcommand{\suzu}[1]{\textcolor{black}{#1}}

\def \bmk{{\bm{k}}}

\begin{document}

\title{
\suzum{Plaquette chirality patterns for robust zero-gap states in $\alpha$-type organic conductor}
}
\author{
Fr\'ed\'eric \textsc{Pi\'echon}$^{1}$,
Yoshikazu \textsc{Suzumura}$^{2}$ 
and
Takao Morinari$^{3}$
}
\address{
$^1$ Laboratoire de Physique des Solides, CNRS UMR 8502, Universit\'e Paris-Sud, F-91405 Orsay Cedex, France \\
$^2$ Department of Physics, Nagoya University, Chikusa-ku, Nagoya 464-8602, Japan \\
$^{3}$ Graduate School of Human and Enviroment Studies, Kyoto University, Kyoto 606-8501, Japan 
}

\ead{suzumura@s.phys.nagoya-u.ac.jp}
\begin{abstract}
 Dirac electrons with a zero-gap state (ZGS) in organic conductor $\alpha$-(BEDT-TTF)$_2$I$_3$
 result from a fine tuning of the seven nearest neighbors transfer integrals ($a_1, a_2, a_3, b_1, b_2, b_3, b_4$) between the four molecules of the unit cell.
In this work 
we show that for given moduli $|a_1|,...|b_4|$, 
the possibility of having Dirac electron with a ZGS at $3/4$ (or $1/4$) filling
strongly depends on the specific configurations of signs of the seven transfer integral.
More precisely it is possible to classify the sign configurations into essentially four classes determined by
\suzum{$\chi_{a}={\rm sign} (a_2a_3)$} and $\chi_b={\rm sign} (b_1b_2b_3b_4)$. 
Using extended numerics, we show that for \suzum{ both weak and  large}  inhomogeneity in the moduli, 
the class $(\chi_a,\chi_b)=(-,-)$ is the most favorable to find Dirac electrons with ZGS at $3/4$ (or $1/4$) filling.
 For the class $(\chi_a,\chi_b)=(+,+)$ \suzum{in the opposite case,} 
  we never found any ZGS at either $1/4$ or $3/4$ filling.
The last two classes \suzum{given by} $(\chi_a,\chi_b)=(+,-)$ and $(\chi_a,\chi_b)=(-,+)$ 
 \suzum{corresponding to}  an intermediate situation; 
they allow for ZGS at $3/4$ (resp. $1/4$) filling but are much less favorable
than class \suzu{$(\chi_a,\chi_b)=(-,-)$}. 
As a matter of fact, all previous numerical studies of Dirac electrons and ZGS in $\alpha$-(BEDT-TTF)$_2$I$_3$ correspond to class $(\chi_a,\chi_b)=(-,+)$. 

\end{abstract}





\section{Introduction} 

It is now well established that organic conductors  $\alpha$-(BEDT-TTF)$_2$I$_3$ \cite{Mori1984_CL},  
exhibit a zero gap-state (ZGS) characteristic of the presence of  massless Dirac electrons close to the Fermi energy \cite{Katayama2006_JPSJ75}.
However, compared to graphene, the Dirac cones are strongly tilted resulting in a strong anisotropy of the Dirac electron velocity components.
Furthermore, the fact that $\alpha$-(BEDT-TTF)$_2$I$_3$ has a unit cell composed of four inequivalent molecules 
leads to some unique physical properties associated to this ZGS. In particular,
as revealed by recent NMR measurements, the anisotropy of the Dirac particles appears strongly related 
to their local spectral weight on the four inequivalent molecules per unit cell; 
such that the  smallest (largest) velocity Dirac particles have their largest weight 
on the electron rich (poor) molecule site \cite{Tajima2009_STAM10,Kobayashi2009_STAM10,ZGS_review_2014}.

\begin{figure}
  \centering
\includegraphics[width=7cm]{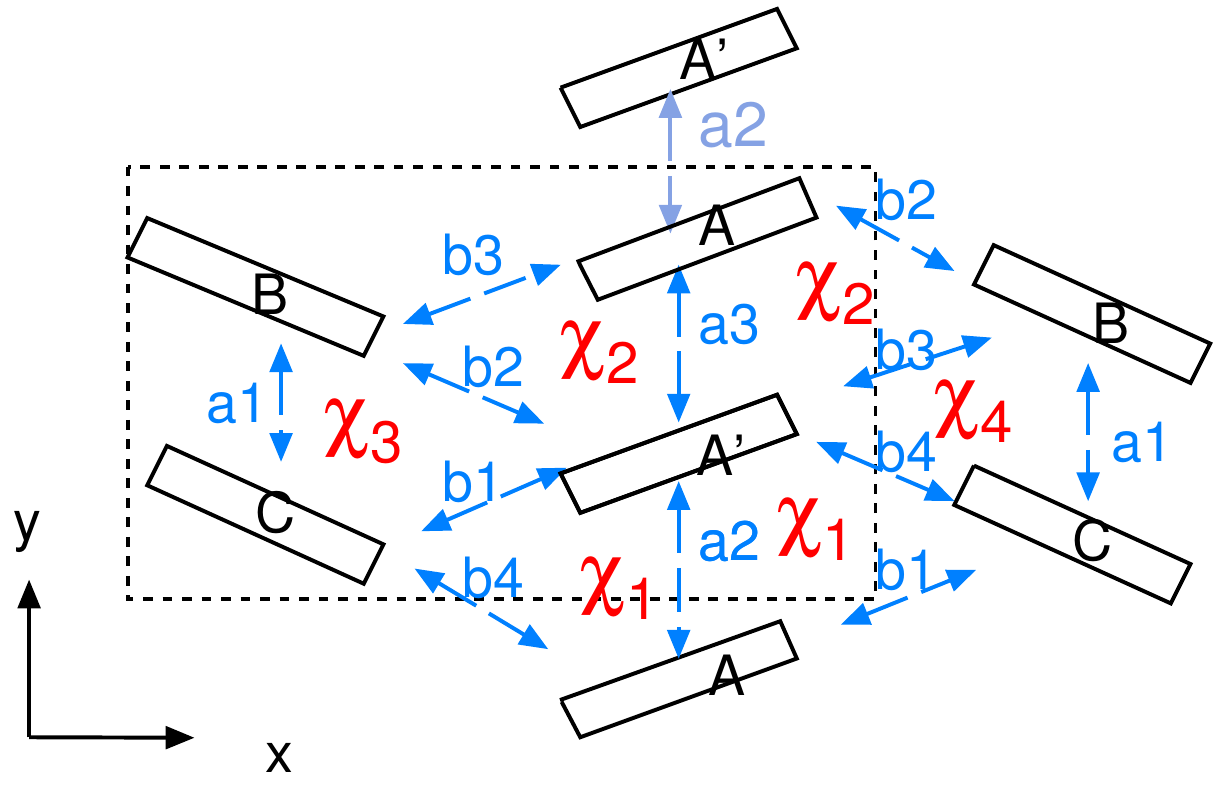}   
  \caption{(Color online)
Structure  
 of $\alpha$-(BEDT-TTF)$_2$I$_3$,
 with four molecules   A, A', B, and C per unit cell (dotted square). 
Inversion centers are located on sites B, C, and 
the middle point between A and A'.\cite{Mori1984_CL}.
The seven nearest-neighbor transfer integrals are shown by  $ a_1 , \cdots, b_4$.  
The chirality of four distinct triangular plaquettes is shown by $\chi_1$, $\chi_2$,  $\chi_3$, and $\chi_4$.
}
\label{fig:structure}
\end{figure}

From a theoretical point of view, the electronic properties of $\alpha$-(BEDT-TTF)$_2$I$_3$ are usually described by
a tight-binding model with seven nearest-neighbor transfer integrals  $a_1, \cdots, b_4$ between the four molecules  
A, A', B, and C of the unit cell. As illustrated on figure Fig.~\ref{fig:structure}, this model presents an inversion symmetry
with inversion centers located on B,C and the middle point between A and A'.\cite{Mori1984_CL}.
Thank's to this inversion symmetry, recent studies have used Fu-Kane topological argument 
in order to determine quantitatively the existence domain of Dirac electrons from the knowledge of the 
energy and inversion parity eigenvalues at the four time reversal invariant momenta (TRIM) \suzum{\cite{Fu2007_PRB76,Piechon2013_JPSJ,Mori2013_JPSJ}}. 
Despite their usefulness, these studies do not allow to distinguish
 ZGS systems in which Dirac electrons are at the Fermi energy  (with a vanishing density of states at the Fermi energy)
from metallic systems in which there is an indirect overlapp between valence 
and conduction band such that the Dirac electrons are below (or above) the Fermi energy 
\suzum{\cite{Mori2010_JPSJ,Suzumura2013_JPSJ}}.

The purpose of the present work is to clarify the relative role played by the 
 the inhomogeneity in the moduli and in the signs of the seven transfer integrals 
on the appearance and stability of Dirac electrons with ZGS.
In fact as explained in more details below, for given moduli $|a_1|,...|b_4|$, 
the possibility of having Dirac electron with a ZGS at $3/4$ (or $1/4$) filling
strongly depends on the specific configurations of signs of the seven transfer integral.
More precisely it is possible to classify the sign configurations into essentially four classes determined by
\suzum{$\chi_{a}={\rm sign} (a_2a_3)$} and $\chi_b={\rm sign} (b_1b_2b_3b_4)$. 
Using extensive numerics, we show that for weak, large and no inhomogeneity in the moduli, the
class $(\chi_a,\chi_b)=(-,-)$ is most favorable to find Dirac electrons with ZGS at $3/4$ (or $1/4$) filling.
The classes  $(\chi_a,\chi_b)=(+,-)$ and $(\chi_a,\chi_b)=(-,+)$ constitute equivalent but less favorable situation 
for the occurrence of Dirac electrons with ZGS  at $3/4$ (or $1/4$) filling. Finally,
the last class is $(\chi_a,\chi_b)=(+,+)$ and for the range of inhomogeneity that was explored no ZGS was obtained for either $1/4$ or $3/4$ filling. 

The paper is organized as follows. In \S 2 we present the model and the main quantities of interest. 
In \S 3 we first introduce the chirality patterns and their properties and then using properties established in Appendix A we explained the relation
between the chirality patterns and the energy band structure and local density of states.
In \S 4 we present our extensive numerical studies of the existence of ZGS at $3/4$ filling for the three pertinent classes of chirality patterns.
In \S 5 we summarized our main results.

\section{Tight-binding model}

The starting point is the $4 \times 4$ Bloch Hamiltonian matrix $H(\bm{k})$ 
corresponding to the real space tight-binding model shown in Fig.~\ref{fig:structure}:
\begin{equation}
H(\bm{k})= 
\begin{pmatrix}
0 & a&b&c\\
a^*&0 &b^* e^{i k_x} & c^* e^{i k_x+i k_y}\\
b^* & b e^{-i k_x} & 0 &f\\
c^* & c e^{-i k_x-ik_y} &{f}^*&0
\end{pmatrix} \ , 
\label{eq:H}
\end{equation}
with $a=a_3 + a_2 {\rm e}^{i k_y}$, $b = b_3 + b_2 {\rm e}^{i k_x}$,   
$c={\rm e}^{ i k_y}(b_4+b_1 {\rm e}^{i k_x})$ and 
$f=a_1 (1+ {\rm e}^{i k_y})$ 
(the lattice constant is taken as unity).
The Boch basis $|j\bm{k}\rangle$ ($j=A,A',B,C$) was chosen such that
the Hamiltonian matrix is Bloch periodic $H({\bmk}+\bm{G})=H({\bmk})$ with ${\bm{G}}$ a reciprocal lattice vector.
In this Bloch basis, the energy eigenstate $|E_n(\bm{k})\rangle$ of the $n^{th}$ band is represented
by a four components vector $u_n(\bm{k})^T\equiv (u_n^A,u_n^{A'},u_n^B,u_n^C)$ such that
$H(\bm{k}) u_n(\bm{k}) = E_n(\bm{k}) u_n(\bm{k})$ where by convention
$E_1(\bm{k}) < E_2(\bm{k}) < E_3(\bm{k}) < E_4(\bm{k})$.
The local density of states (LDOS) on each site is then obtained as ($j=A,A',B,C$)
\begin{equation}
D_{j}(E)=\int \frac{d\bm{k}}{4\pi^2} \sum_n |u_n^j(\bm{k})|^2\delta(E-E_n(\bm{k})).
\end{equation}
For $3/4$ filling, the existence of a pair of Dirac points at $\pm \bm{k}_D$ implies $E_4(\bm{k}_D) = E_3(\bm{k}_D)=E_{\bm{k}_D}$  
with $\bm{k}_D \not= \bm{G}/2$; the existence of a ZGS further necessitates that $E_4(\bm{k}) > E_{\bm{k}_D} > E_3(\bm{k})$  for any $\bm{k}$.
In that situation the LDOS verify $D_{j}(E) \simeq \frac{|E-E_D|}{v_j^2}$ for $E$ near $E_D$ where $v_j$ 
is the site dependent angular averaged Fermi velocity.

\section{Classification in terms of chirality patterns}

For fixed moduli ($|a_1|, \cdots |b_4|$), the energy bands $E_n(\bm{k})$ and the eigenfunction components $u_n^j(\bm{k})$ 
have the freedom of the sign of the seven hopping parameters, as a result  one may expected to obtain $2^7$ distinct LDOS $D_{j}(E)$.
As explained in \ref{ldos}, the number of distinct LDOS $D_{j}(E)$ 
is entirely determined by the number of patterns of the four chiralities $\chi_{i}=\pm$ ($i=1,2,3,4$) \suzum{\cite{Piechon2014}}
\begin{equation}
 \chi_1 = {\rm sign} (a_2 b_1 b_4), \ \ \   \chi_2 = {\rm sign} (a_3 b_2 b_3), \ \ \ \chi_3 = {\rm sign} (a_1 b_1 b_2), \ \ \ \chi_4 = {\rm sign} (a_1 b_3 b_4),
\end{equation}
where each chirality $\chi_{i}$ encodes the sign of the product of the transfer integrals for each 
of the four distinct triangular plaquette that compose the unit cell.
The are in total $2^4=16$  chirality patterns $(\chi_1,\chi_2,\chi_3,\chi_4)$ which are summarized in Table \ref{table_1}.

\begin{figure}
  \centering
  \begin{tabular}{ccc}
 \Large $A_3$&\Large$A_7$&\Large $S_6$\\
{\includegraphics[width=4.cm]{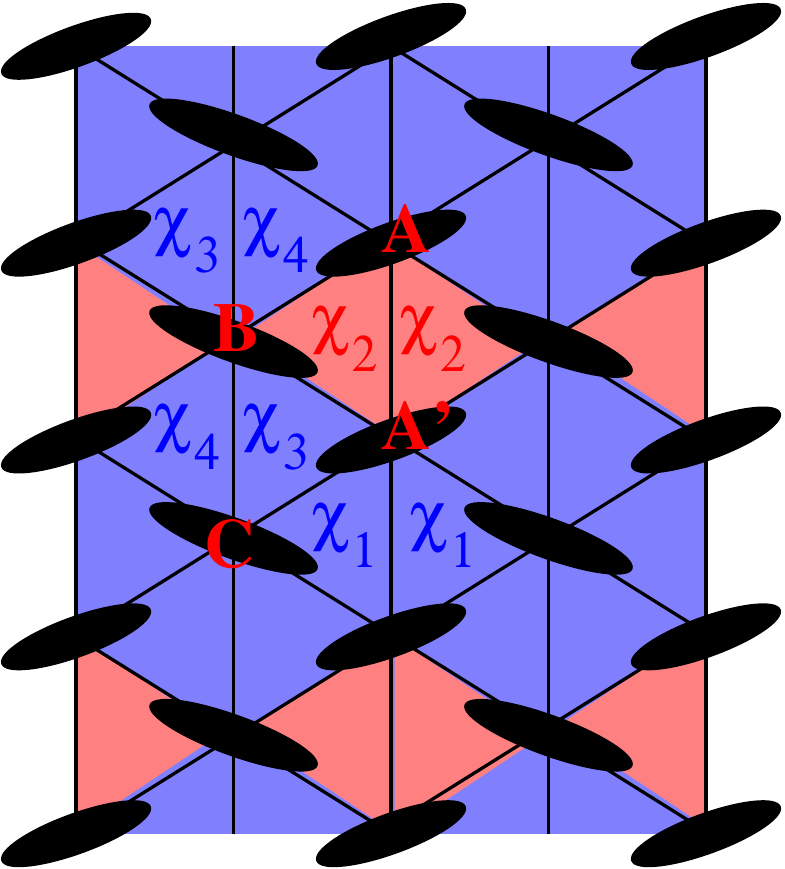}}&
{ \includegraphics[width=4.cm]{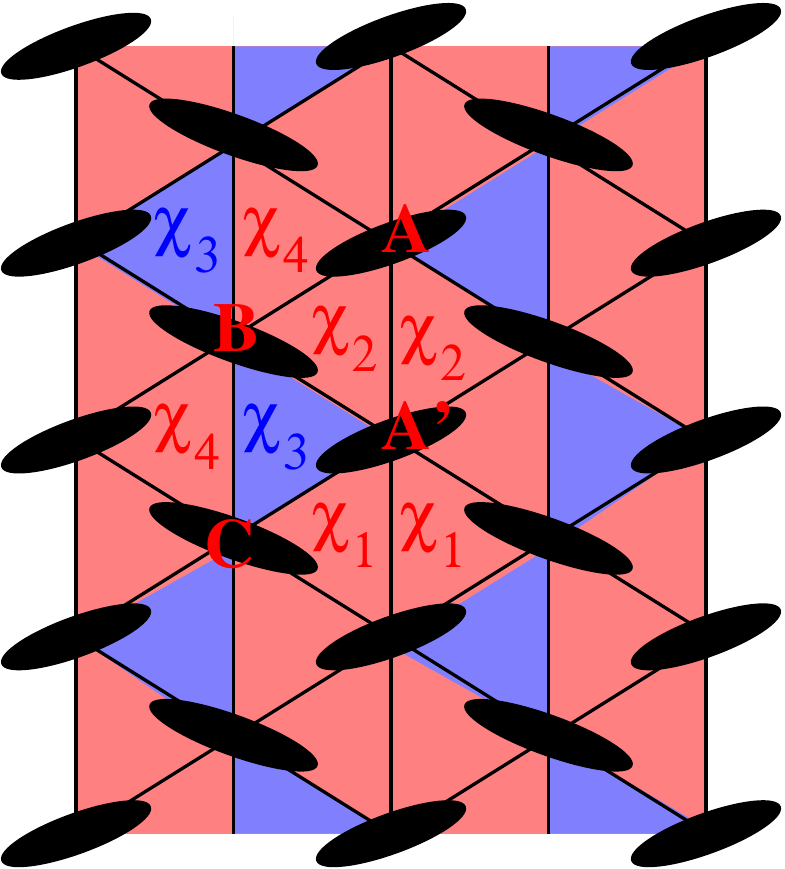}}&
{\includegraphics[width=4.cm]{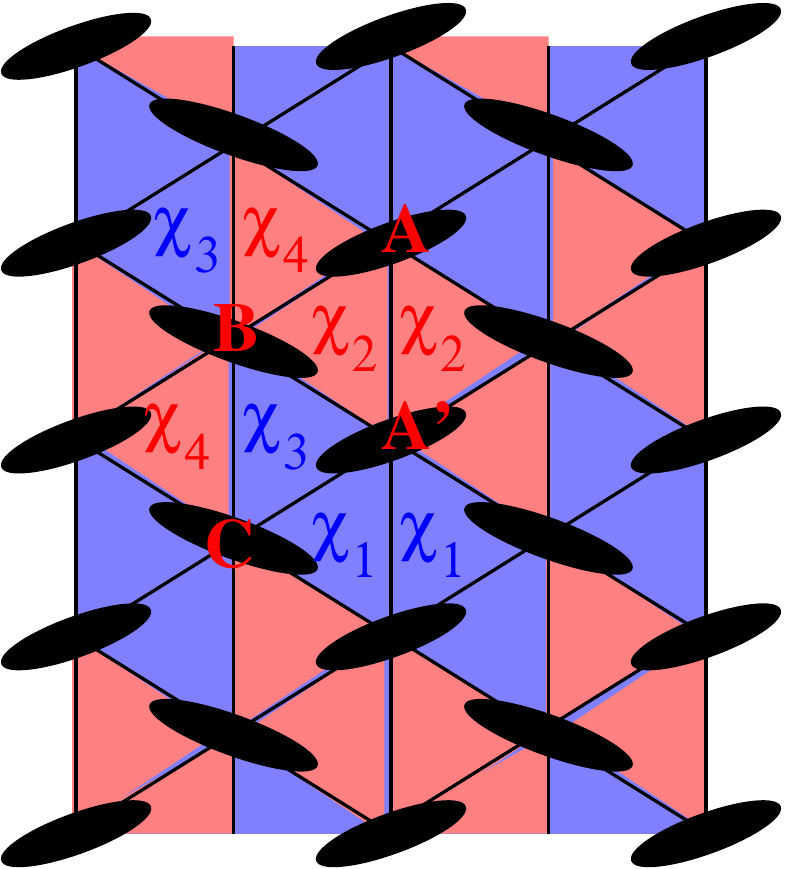}}\\
\Large $S_2$&\Large $S_1$&\\
{\includegraphics[width=4.cm]{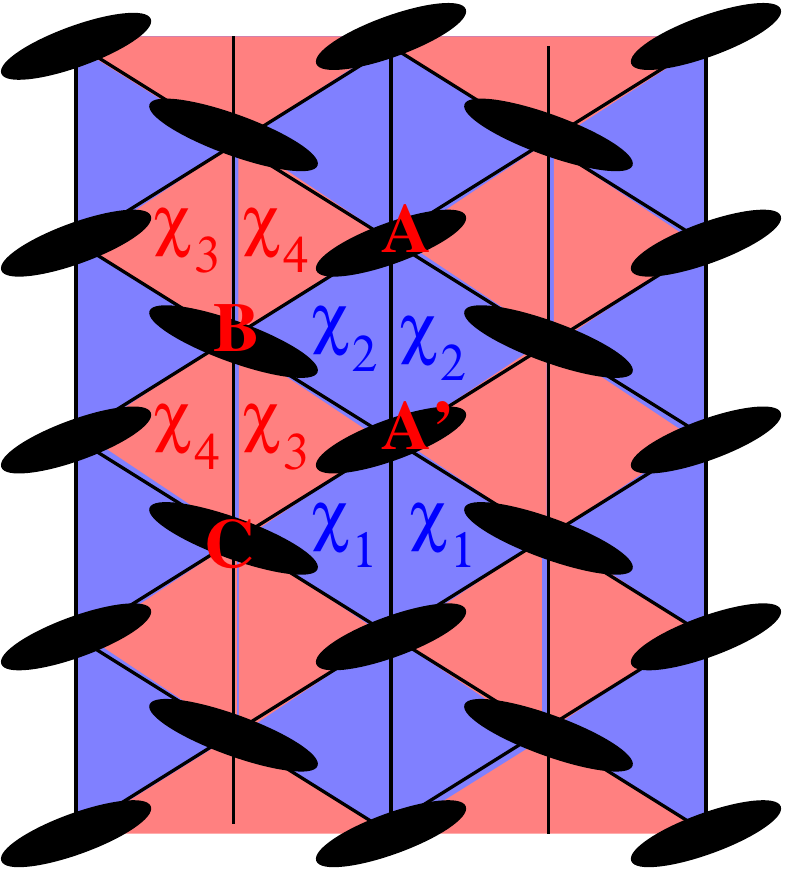}}&
{\includegraphics[width=5.cm]{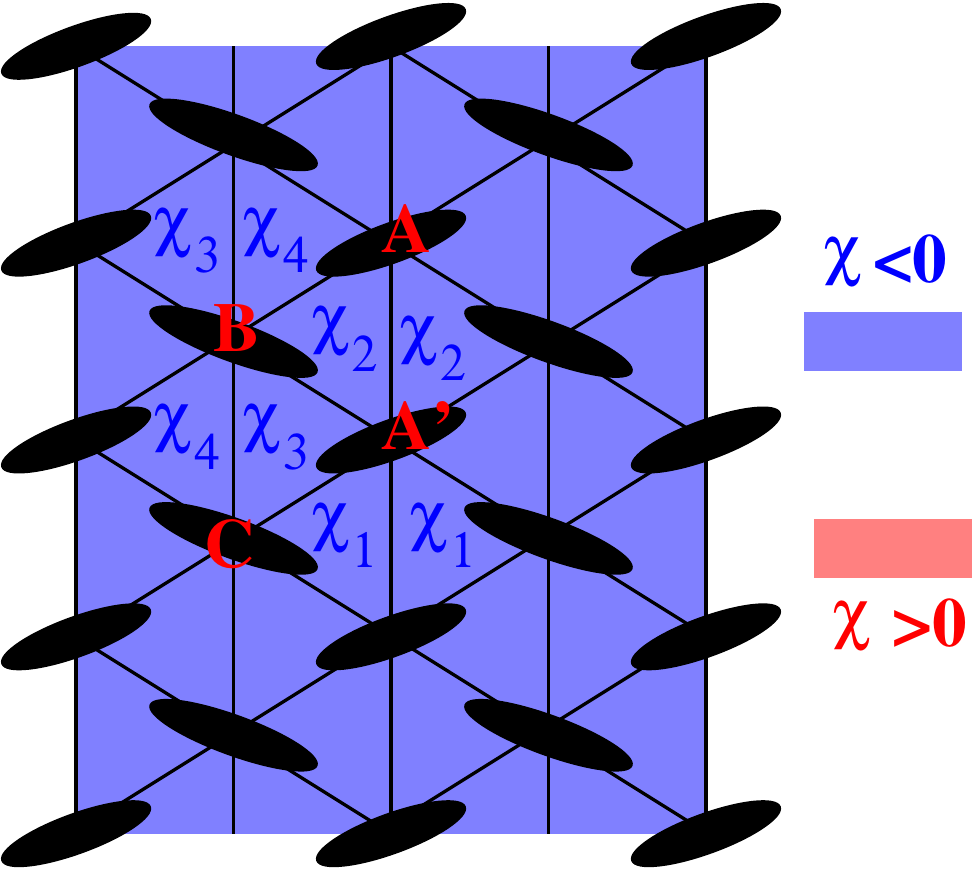}}  &
 \end{tabular}     
  \caption{(Color online)
Real space representation of the five distinct triangular plaquette chirality patterns 
$A_7$, $A_3$, $S_6$ , $S_1$ , $S_2$.
Triangular plaquette with $\chi_i=+$ ($\chi_i=-$) are in red (blue).
Chirality conjugaison $C_{\chi}$ amounts to switching the color of each plaquette, in that way
one obtains the corresponding five patterns $A_2$, $A_6$, $S_5$ , $S_8$ , $S_7$.
By applying $M_{a}$ the five patterns $A_7$, $A_3$, $S_6$ , $S_1$ , $S_2$ transform into respectively $A_8$, $A_3$, $S_5$ , $S_1$ , $S_2$.
Similarly, by applying $T_{a}$ they transform into respectively $A_8$, $A_4$, $S_3$ , $S_1$ , $S_2$.
}
\label{patterns}
\end{figure}

Interestingly it appears that the chirality patterns are not all {\em independent}, more precisely one can essentially define
five classes of patterns with independent real space {\em topology}: ${\mathscr C}_1=(A_1,A_2,A_7,A_8)$, ${\mathscr C}_2=(A_3,A_4,A_5,A_6)$, 
${\mathscr C}_3=(S_3,S_4,S_5,S_6)$, ${\mathscr C}_{4a}=(S_1,S_8)$, ${\mathscr C}_{4b}=(S_2,S_7)$.
As an illustration of their distinct {\em topology}, the chirality patterns associated to $A_3,A_7,S_6,S_2,S_1$ are represented on Fig. \ref{patterns}(a),..(e).
Each class is characterized by the fact that it is globally invariant under the following three transformations: 
(i) chirality conjugaison $C_{\chi}$: $(\chi_1,\chi_2,\chi_3,\chi_4)\rightarrow (-\chi_1,-\chi_2,-\chi_3,-\chi_4)$, 
(ii) mirror along y-axis chains $M_y$: $(\chi_1,\chi_2,\chi_3,\chi_4)\rightarrow (\chi_1,\chi_2,\chi_4,\chi_3)$, 
(iii) translation  by half unit along y-axis $T_y$: $(\chi_1,\chi_2,\chi_3,\chi_4)\rightarrow (\chi_2,\chi_1,\chi_4,\chi_3)$.
As example one easily verifies the identities $A_8=T_y [A_7]=M_y [A_7]$, $A_1=C_{\chi}T_y [A_7]$ and $A_2=C_{\chi} [A_7]$. 
Note that for patterns $S_i$ in classes ${\mathscr C}_{4a},{\mathscr C}_{4b}$ one has $S_i=M_y [S_i]=T_y [S_i]$; 
in that sens it is more meaningful to define a single class ${\mathscr C}_{4}=(S_1,S_8,S_2,S_7)$. 
In fact it then appears that the four classes ${\mathscr C}_{i}$ are equivalent to classify 
the $16$ chirality patterns in terms of the two effective chiralities 
\begin{equation}
 \chi_a=\chi_1\chi_2\chi_3\chi_4=\rm{sign}(a_2a_3), \ \ \ \ \ \chi_b=\chi_3\chi_4=\rm{sign}(b_1 b_2 b_3 b_4),
\end{equation}
such that
${\mathscr C}_{i}\equiv (\chi_a,\chi_b)$ with ${\mathscr C}_{1}=(-,-)$, ${\mathscr C}_{2}=(-,+)$, ${\mathscr C}_{2}=(+,-)$ and ${\mathscr C}_{4}=(+,+)$.

We terminate this section by listing other properties relating chirality patterns and LDOS.\\
$P_1$: Since there are only $2^4$ patterns, it implies that there are $2^3$ (out of $2^7$) configurations of signs of the seven transfer integrals 
that leads to the same pattern $(\chi_1,\chi_2,\chi_3,\chi_4)$ and to the same LDOS $D_{j}(E)$ (see \ref{appendix_a}).
In other words, without loss of generality, it is always possible to fix arbitrarily the sign of three out of the seven transfer integrals and
hereafter the convention $(b_1,b_2,b_3)>0$ is adopted in all numerical computations.\\
$P_2$: From the properties of the moments of the LDOS (\ref{ldos}), it is straightforward to show that 
chirality conjugation $C_{\chi}$ is associated 
to $D_{j}(E)\rightarrow D_{j}(-E)$.  More physically it means that if a pattern exhibits a ZGS at $1/4$ (resp. $1/2$, $3/4$) filling 
then its chirality conjugate pattern exhibits a ZGS at $3/4$ (resp. $1/2$, $1/4$) filling.\\
$P_3$: In the cases ($|a_2|=|a_3|, |b_1|=|b_3|, |b_2|=|b_4|$), using eq.(\ref{A6}) 
it easily shown that for any pattern $(\chi_1,\chi_2,\chi_3,\chi_4)$, its $M_y$ (or $T_y$) transformed pattern has the same DOS.

As a consequence of properties ($P_2,P_3$), for the cases ($|a_2|=|a_3|, |b_1|=|b_3|, |b_2|=|b_4|$) it is thus sufficient to study the ZGS and LDOS properties 
of patterns  $A_3,A_7,S_6,S_2,S_1$ to deduce the properties of all other patterns.  
Lastly, we note that all previous numerical studies of Dirac points and ZGS in $\alpha$-(BEDT-TTF)$_2$I$_3$
corresponds to the pattern $A_3$.

\begin{table}
\caption{Chirality patterns $(\chi_1,\chi_2,\chi_3,\chi_4)$.
The group $A_{i}$ is associated to $\chi_1\chi_2\chi_3\chi_4=-$ 
whereas the group $S_{i}$ is associated to $\chi_1\chi_2\chi_3\chi_4=+$.}
\begin{center}
\begin{tabular}{cccccccccccccccccc}
\hline\noalign{\smallskip}
   & $A_1$ & $A_2$& $A_3$& $A_4$ & $A_5$ & $A_6$ & $A_7$ & $A_8$ 
 & $S_1$ & $S_2$& $S_3$& $S_4$ & $S_5$ & $S_6$ & $S_7$ & $S_8$ \\
\noalign{\smallskip}\hline\noalign{\smallskip}
$\chi_1$ & - & - & - & - & + & + & + & + &-&-&-&-&+&+&+&+\\
$\chi_2$ & - & - & + & + & - & - & + & + &-&-&+&+&-&-&+&+\\
$\chi_3$ & - & + & - & + & - & + & - & + &-&+&-&+&-&+&-&+ \\
$\chi_4$ & + & - & - & + & - & + & + &  - &-&+&+&-&+&-&-&+\\
\noalign{\smallskip}\hline
\end{tabular}
\end{center}
\label{table_1}
\end{table}

\section{Existence and stability of ZGS at $3/4$ fillings}

In the following, we concentrate our numerics on the study of ZGS at $3/4$ filling for patterns $A_7,A_3,S_6$ of classes
${\mathscr C}_{1},{\mathscr C}_{2},{\mathscr C}_{3}$ respectively. 
We stress that patterns $A_8,A_4,S_5$ present very similar properties as $A_7,A_3,S_6$.
We discard patterns of class ${\mathscr C}_{4}$ because
for the range of parameters that we have studied  we never found ZGS at either $1/4$ or $3/4$ filling for these patterns.

\subsection{ZGS and LDOS properties for homogeneous modulus $|a_1|=|a_2|=....=|b_4|=1$}

\begin{figure}
  \centering
  \includegraphics[]{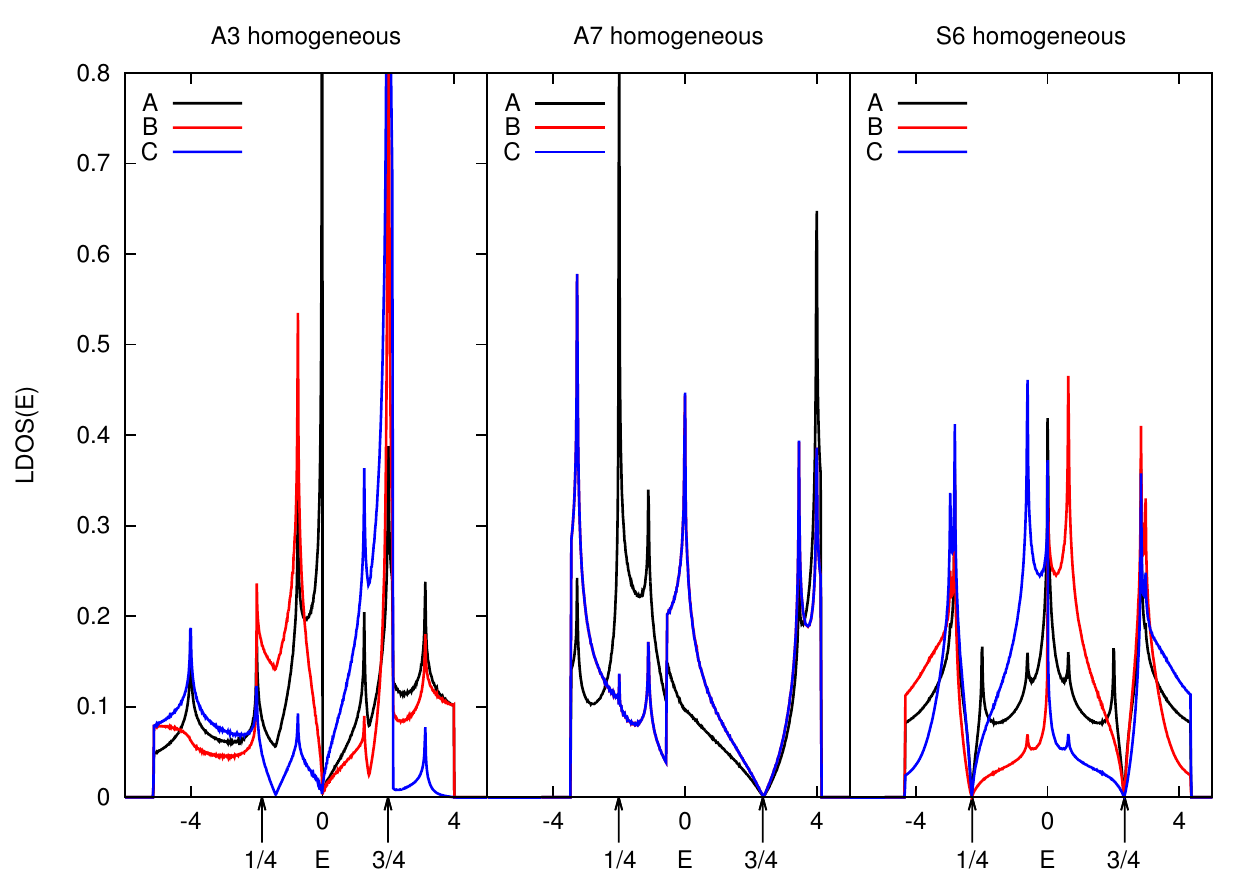}
      
  \caption{(Color online)
Local densities of states for patterns $A_3,A_7,S_6$ in the case of homogeneous modulus $|a_1|=|a_2|=....=|b_4|=1$,  
$D_A(E)=D_{A'}(E)$ black line, $D_B(E)$ red line and $D_C(E)$ blue line.
}
\label{ldoshomo}
\end{figure}

As a first illustration of the importance of the chirality pattern on the appearance of ZGS, Fig. \ref{ldoshomo} represents respectively the LDOS
$D_{j}(E)$ ($j=A,B,C$) for patterns $A_3,A_7,S_6$ in the case of homogeneous moduli $|a_1|=|a_2|=....=|b_4|=1$.
On Fig. \ref{ldoshomo}, for each pattern, the energies corresponding to $1/4$ and $3/4$ fillings are indicated by vertical arrows.

For pattern $A_3$, the $A,B,C$ sites have strongly different LDOS; in addition 
at $1/4$ and $3/4$ fillings the system is metallic, moreover there is a strong 1D like VanHove 
singularity for $D_{A,B,C}(E)$ at $3/4$ filling. For pattern $A_7$, the LDOS verify 
$D_B(E)=D_C(E)\ne D_A(E)$; in addition there is a {\em wide} ZGS at $3/4$ filling and a metallic state at $1/4$ filling with a strong VanHove 
singularity for $D_A(E)$.
For pattern $S_6$, the LDOS verify $D_B(E)=D_C(-E)\ne D_A(E)$ moreover there is a {\em narrow} ZGS both for $1/4$ and $3/4$ fillings whereas 
there is a VanHove singularity at $1/2$ filling.

From these results, one can already anticipate that pattern $A_3$ which shows a VanHove singulary at $3/4$ filling may thus require 
some peculiar and probably strong anisotropy in the moduli in order to exhibit a ZGS state at this filling. In a similar manner, one can anticipate
that the ZGS at $3/4$ filling obtained for patterns $A_7$ and $S_6$ may have distinct robustness against the introduction of an anisotropy in the moduli.
More precisely the ZGS of pattern $A_7$ which exhibits a linear LDOS over wide energy range is anticipated to be more robust than the ZGS of pattern $S_6$ 
which is relatively narrow in energy. 

\subsection{ZGS and LDOS properties for BEDT moduli at pressure $P=6$Kbars}

\begin{figure}
\centering
\includegraphics[]{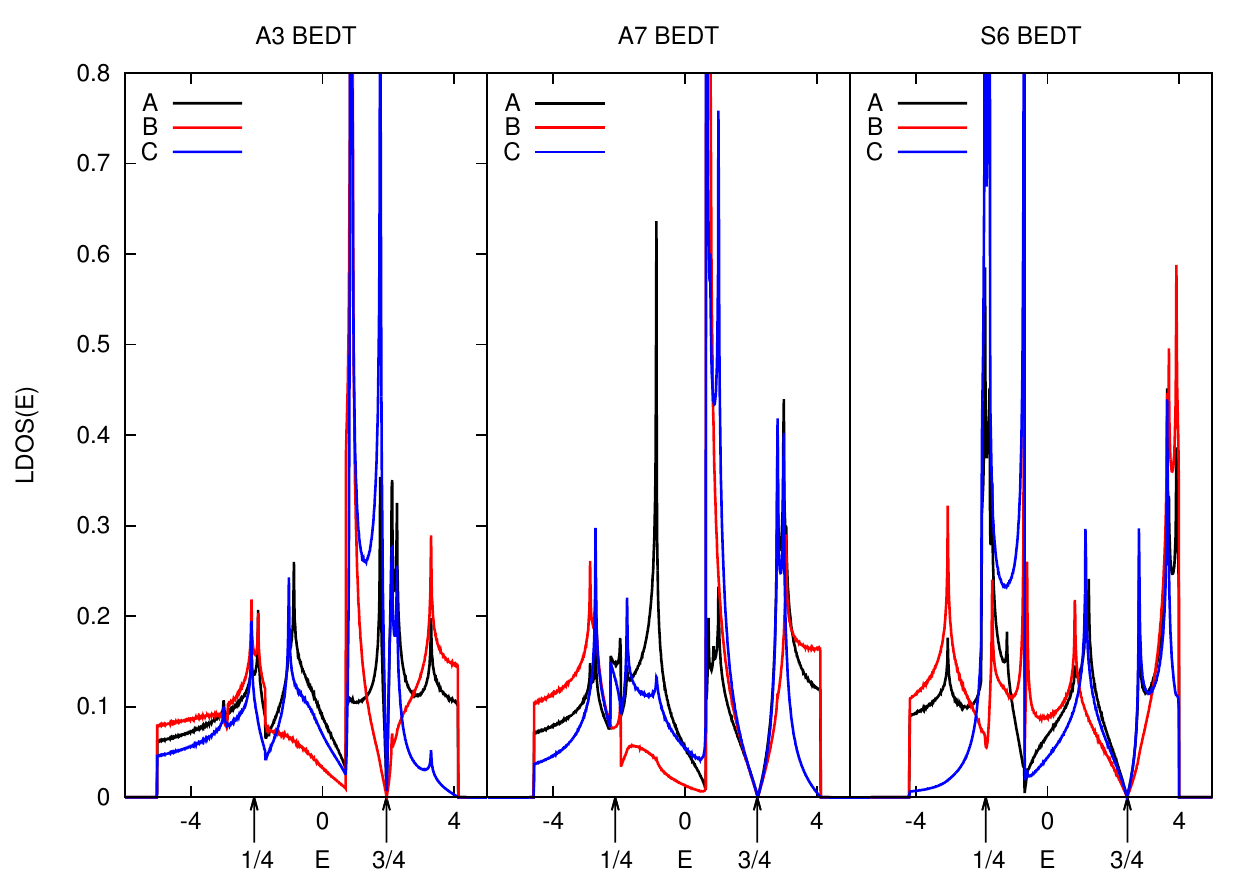}
      
\caption{(Color online) Local densities of states for patterns $A_3,A_7,S_6$ in the case of BEDT moduli at pressure $P=6$ Kbars 
($|a_2|$,$|a_3|$,$|b_1|$,$|b_2|$,$|b_3|$,$|b_4|$) =$|a_1|(2.23,0.39,2.86,3.46,1.70,0.58)$   \cite{Kondo2003,Kondo2009}.
$D_A(E)=D_{A'}(E)$ black line, $D_B(E)$ red line and $D_C(E)$ blue line.
}
\label{ldosbedt}
\end{figure}

As a first illustration of the combined role of chirality pattern and inhomogeneity in the moduli,  Fig \ref{ldosbedt} represents the LDOS
$D_{j}(E)$ ($j=A,B,C$) for patterns $A_3,A_7,S_6$ in the case of BEDT moduli at pressure $P=6$ Kbars 
($|a_2|$,$|a_3|$,$|b_1|$,$|b_2|$,$|b_3|$,$|b_4|$) =$|a_1|(2.23,0.39,2.86,3.46,1.70,0.58)$\cite{Kondo2003,Kondo2009},
for which it is known that pattern $A_3$ exhibits a ZGS at $3/4$ filling.
The value of $|a_1|$ is taken such that the second moment of the DOS is the same as for homogeneous modulus.

At first sight, the main consequence of the strong inhomogeneity in the moduli is that the shape of the LDOS are now qualitatively 
much more similar for all patterns. In particular, now all patterns exhibit a ZGS at $3/4$ filling, however for pattern $A_3$ the ZGS is very narrow 
in energy and with strong VanHove singularities very near the Dirac point indicating a strong tilt of the
Dirac cones. Note that for pattern $S_6$ there is no more a ZGS at $1/4$ filling where it is now metallic 
with a strong VanHove singularity. Overall, these results seem to indicate that chirality patterns might play a weaker role for strong inhomogeneity.

\subsection{ZGS properties for inhomogeneous moduli}

The combined roles of chirality pattern and  inhomogeneity of moduli 
on the existence and stability of ZGS at $3/4$ fillings are now examined more systematically for the three patterns $A_3,A_7,S_6$.
To characterize the inhomogeneity of the moduli it is convenient to define the following six dimensionless parameters:
\begin{equation}
\begin{array}{lll}
\Delta_a=\frac{|a_2|-|a_3|}{|a_2|+|a_3|},& \Delta_b=\frac{|b_1|-|b_3|}{|b_1|+|b_3|},& \Delta_{b'}=\frac{|b_2|-|b_4|}{|b_2|+|b_4|},\\
z_a=\frac{|a_2|+|a_3|}{2|a_1|},&  z_b=\frac{|b_1|+|b_3|}{2|a_1|},& z_{b'}=\frac{|b_2|+|b_4|}{2|a_1|}.
\end{array}
\end{equation}
The quantities $-1\le \Delta_{a,b,b'}\le 1$ characterize the {\em dimerization} along the chains (AA') and the {\em trimerization} 
along the two effective chains (BACA'). The ratio $z_a,z_b,z_{b'}$ quantify essentially the relative anisotropy 
between the different chains as compared to chain (BC).
Within this parametrization, the case of homogeneous moduli corresponds to $\Delta_{a,b,b'}=0$ and $z_{a,b,b'}=1$, whereas
the case of BEDT moduli at $P=6$ corresponds to $(z_a,z_b,z_{b'},\Delta_a,\Delta_{b},\Delta_{b'})=(1.31,2.27,2.02,0.70,0.25,0.71)$.
In the following, for a given set of parameters $(z_{a,b,b'},\Delta_{a,b,b'})$, the value of $a_1$ is
determined so as to keep the second moment of the DOS equal to that of the homogeneous case; in that way the total bandwidth is roughly constant:
$|a_1|=\sqrt{6/[1+z_a^2(1+\Delta_a ^2)+2z_b^2(1+\Delta_b ^2)+2z_{b'}^2(1+\Delta_{b'} ^2)]}$
With the convention $(b_1,b_2,b_3)>0$, the seven hopping parameters read:
\begin{equation}
  \begin{array}{l}
a_1=\chi_3 |a_1|,\\
a_2=\chi_1 \chi_3 \chi_4 z_a (1+\Delta_a) |a_1|,\\
a_3=\chi_2  z_a (1-\Delta_a)|a_1|,\\
b_1= z_b (1+\Delta_b) |a_1|,\\
b_3=z_b (1-\Delta_b) |a_1|,\\
b_2=z_{b'} (1+\Delta_{b'})|a_1|,\\
b_4=\chi_3 \chi_4 z_{b'} (1-\Delta_{b'}) |a_1|,\\
  \end{array}
\end{equation}
For a given chirality pattern and a given set of parameters ($z_{a,b,b'}$,$\Delta_{a,b,b'}$), 
to characterize the existence of a ZGS at $3/4$ filling 
we proceed as follows.  In a first step the energy eigenvalues $E_n({\bf k})$ are computed over 
a regular mesh of $400 \times 400$ k-points in the BZ such that the maximum $E_n ^{\rm{max}}({\bf k}_n^{\rm{max}})$ 
and minimum $E_n ^{\rm{min}}({\bf k}_n^{\rm{min}})$ of each band are determined. 
In a second step a ZGS at filling $3/4$ is obtained whenever the following three conditions are simultaneously verfied:
$(i)$: ${\bf k}_{4}^{\rm{min}},{\bf k}_3^{\rm{max}}\ne {\bf G/2}$; $(ii)$: $|{\bf k}_{4}^{\rm{min}}-{\bf k}_3^{\rm{max}}| \ll \delta $;
$(iii)$: $0<E_{4} ^{\rm{min}}({\bf k}_{4}^{\rm{min}})-E_{3} ^{\rm{max}}({\bf k}_{3}^{\rm{max}})\le \delta \ll 1$ 
where in practice $\delta \sim (\pi/100)$ was taken. If either $(i)$ or $(ii)$ are not verified or if 
$E_{4} ^{\rm{min}}({\bf k}_{4}^{\rm{min}})-E_{3} ^{\rm{max}}({\bf k}_{3}^{\rm{max}})<0$
 the system is considered as metallic, whereas it is considered as gapped if $E_{4} ^{\rm{min}}({\bf k}_{4}^{\rm{min}})-E_{3} ^{\rm{max}}({\bf k}_{3}^{\rm{max}})>\delta$.
Note that taking a smaller value for $\delta$ will a priori decrease the number of configuration which can be considered as metallic or ZGS. 

For each of the three patterns ($A_3, A_7, S_6$), the numerical determination of possible ZGS at $3/4$ fillings
is done systematically for parameters $z_{a,b,b'} \in (0,0.2,0.4,...,4)$ and for six distinct dimerization configurations $(\Delta_{a}, \Delta_{b},\Delta_{b'})$
as given in table \ref{table_2}; such that configuration I is equivalent to no dimerization, II to weak and VI to strong dimerization. 
For a given pattern and a given  dimerization configuration the existence of ZGS is thus explored for $(21)^3=9261$ 
distinct configurations of parameters $z_{a,b,b'}$.

\begin{table}
\caption{Dimerization configurations $(\Delta_{a}, \Delta_{b},\Delta_{b'})$}
\begin{center}
\begin{tabular}{ccccccc}
\hline\noalign{\smallskip}
   & I & II& III& IV & V& VI  \\
\noalign{\smallskip}\hline\noalign{\smallskip}
$\Delta_{a}$& 0 & 0.25 & 0.25& 0.5& 0.75&0.75\\
$\Delta_{b}$& 0 & 0.25 & 0.75& 0.5& 0.25&0.75\\
$\Delta_{b'}$& 0 & 0.25 & 0.25& 0.5& 0.75&0.75\\
\noalign{\smallskip}\hline
\end{tabular}
\end{center}
\label{table_2}
\end{table}

\subsubsection{histrogram distribution of the number of gapped, ZGS and metallic phases}

\begin{figure}

 \centering
 \includegraphics[]{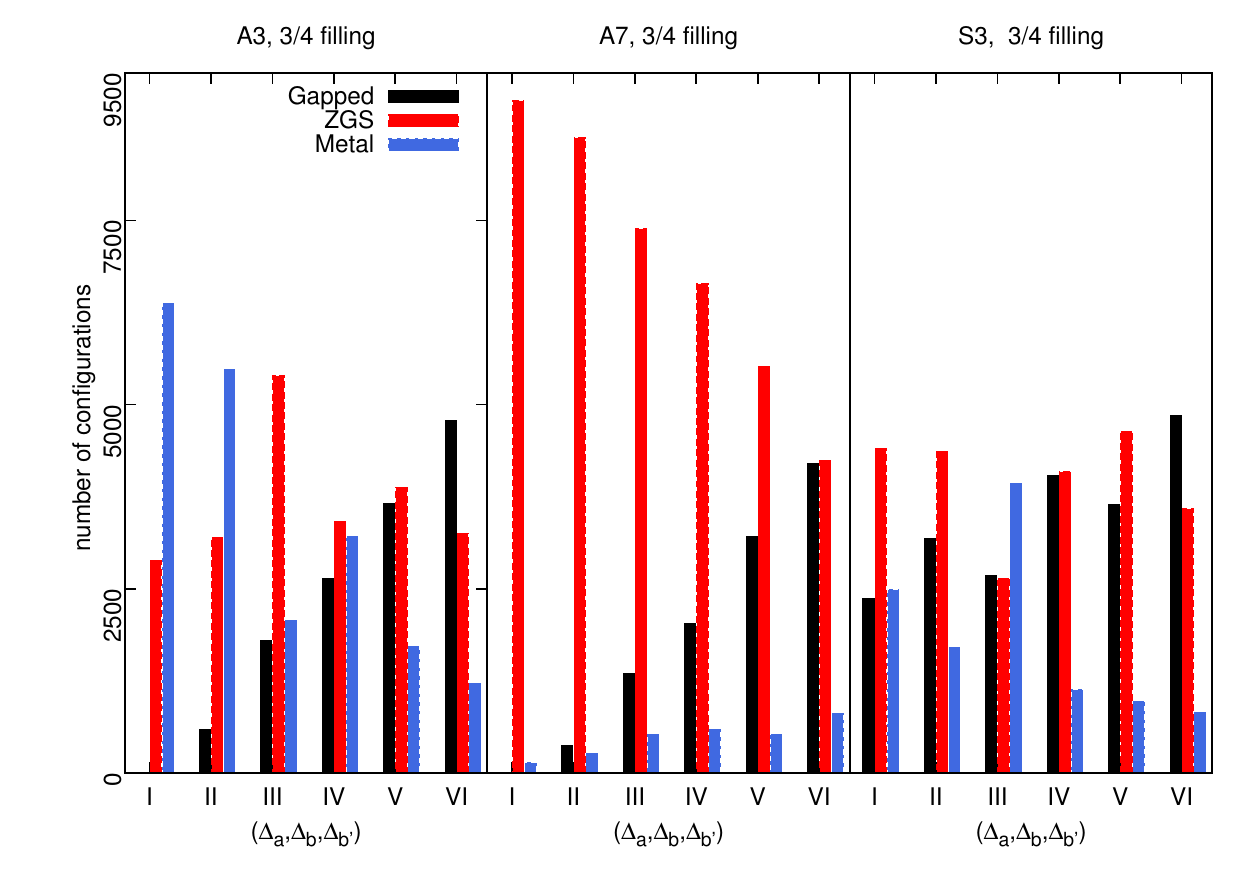}
 \caption{ histogram distribution of the number of configurations that are gapped (black), ZGS (red) or metallic (blue) for patterns $A_3$, $A_7$, $S_6$ at $3/4$ filling
 for the six distinct dimerization configurations (I,II,...VI).}
  \label{histo3}
 \end{figure}

The first statistic Fig.\ref{histo3} represents the histogram distribution of the number of gapped, ZGS and metallic configurations for patterns ($A_3, A_7, S_6$)
at $3/4$ filling and for the six distinct dimerization configurations (I,II,...VI). 

The common and expected feature is that the number of gapped configurations (black bars) increases with the strength of the dimerization along the different chains; 
for all patterns it eventually becomes the dominant phase for strong dimerization (VI). Despite this common feature, we find it relatively surprising that for pattern $S_6$, 
the number of gapped configurations is always relatively large even for low or no dimerization.
As far as the number of metallic configurations (blue bars) is concerned, its behaviour is very different for the three distinct patterns ($A_3, A_7, S_6$). 
For pattern $A_3$, it represents the dominant phase for small dimerization (I) and then it steadily decreases 
with increasing dimerization and it becomes the rarest phase for large dimerization (VI); this type behaviour is somehow as expected.
For pattern $A_7$, whatever the dimerization, the number of metallic configurations is surprisingly always very small in proportion 
such that roughly speaking this pattern exhibits either a ZGS or a gapped phase.
For pattern $S_6$, the number of metallic configurations slowly decreases when dimerization is increased and it is always below that of gapped and ZGS phases 
excepted for dimerization (III) where it represents the dominant phase. Overall, we find it quite remarkable that for most dimerization configurations 
and for the three patterns ($A_3, A_7, S_6$), the number of metallic configurations remains always smaller than the number of ZGS phases.
In fact, regarding the number of ZGS configurations, the three patterns ($A_3, A_7, S_6$) show distinctive behaviours.
For patterns  ($A_3$,$S_6$) the number of ZGS configurations shows no clear tendency as a function of dimerization strength; for each dimerization configuration
it constitutes around $1/3$ of the 9261 anisotropy configurations, slightly higher for $S_6$ and slightly lower for $A_3$.
More remarkably, for pattern $A_7$, even if the number of ZGS configurations decreases steadily when the dimerization strength is increased, 
it still constitutes the largely dominant phase over the entire studied range of dimerization.
In summary, the main message to retain from Fig.\ref{histo3} is that over the range of anisotropy and dimerization that we have considered, 
the pattern $A_7$ is by far the most favorable for the appearance of stable ZGS at $3/4$ filling whereas 
patterns $A_3,S_6$ are more or less similarly favorable. 

\subsubsection{Anisotropy versus dimerization dependency of ZGS properties}

\begin{figure}
 \centering
 \includegraphics[]{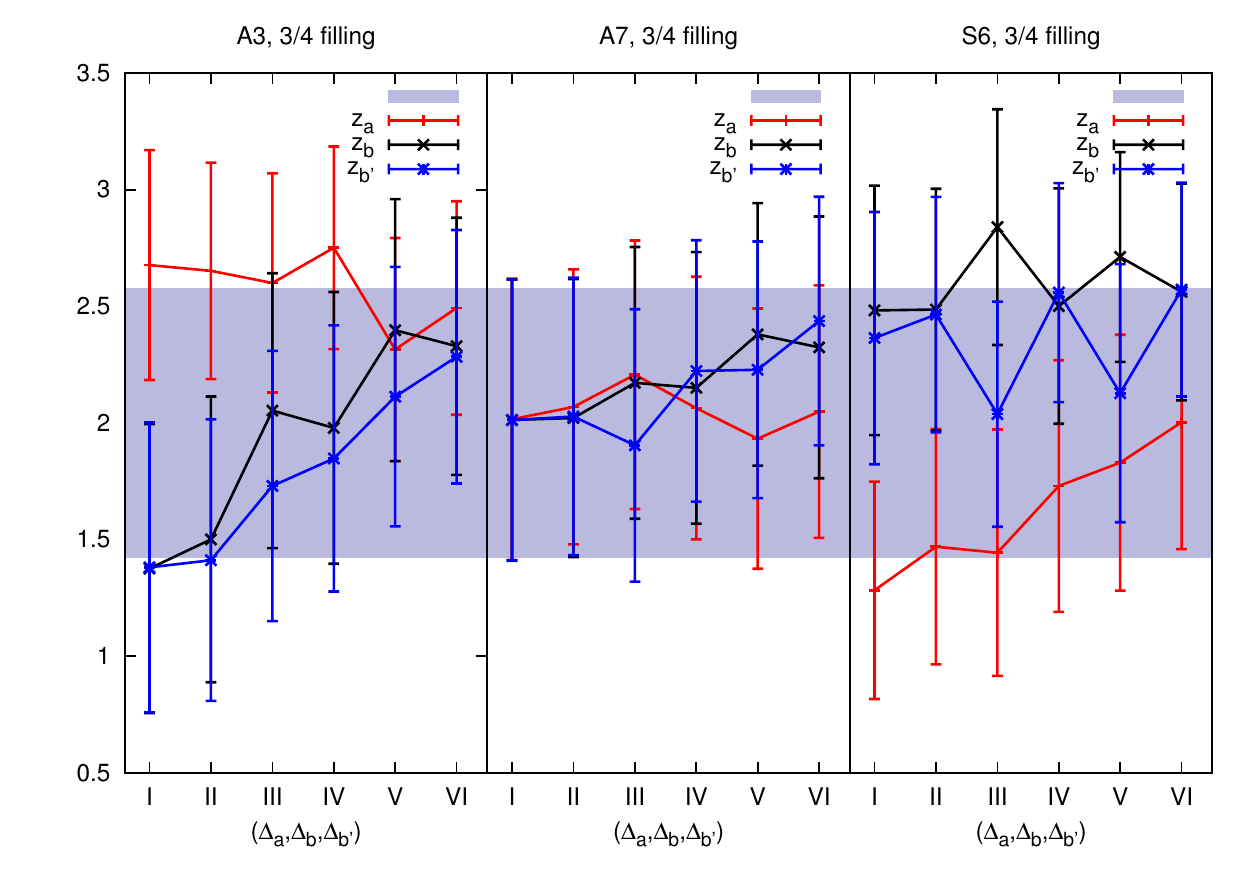}
 \caption{(color online) Averaged anisotropy parameter values $z_a$ (red), $z_b$ (black) $z_{b'}$ (blue) versus dimerization configurations (I,II,...VI) 
 for the ZGS configurations found for patterns $A_3$, $A_7$, $S_3$ at $3/4$ filling. The errorbars encode the meansquare fluctuations. For comparison, the grey region corresponds 
 to the case of a uniform distribution of ZGS configurations over the interval $(z_a,z_b,z_{b'})\in [0,4]$.}
\label{zazbzc3}
 \end{figure}

The previous histograms mainly provide quantitative information of the role played by the dimerization parameters $(\Delta_{a}, \Delta_{b},\Delta_{b'})$ 
on the appearance of ZGS at $3/4$ filling, however they 
do not allow to understand the role played by the anisotropy parameters $z_{a,b,b'}$.
In order to quantify the role played by the anisotropy between the different chains, for each dimerization configuration
we have computed both the average value and the variance of the anisotropy for configurations $z_{a,b,b'}$ that lead to ZGS at $3/4$ filling for patterns $A_3$, $A_7$, $S_6$.
The results are summarized on Fig. \ref{zazbzc3}.
On this figure the grey region corresponds to the case of a uniform distribution of ZGS configurations over the explored interval of anisotropy : $(z_a,z_b,z_{b'})\in [0,4]$;
we reminds that no anisotropy correspond to $z_a=z_b=z_{b'}=1$.
For pattern $A_3$, for no or weak dimerization there is a strong deviation from the uniform distribution, 
more precisely the more favorable anisotropy configurations correspond to large value of $z_a$ and small value of $z_b,z_{b'}$. By increasing the dimerization the average values of 
$z_a,z_b,z_{b'}$ enter the region corresponding to a uniform distribution.
For pattern $S_6$ the behaviour is somewhat symmetric to that of pattern $A_3$, for small dimerization small value of $z_a$ and larger value of $z_b,z_{b'}$ are more favorable to obtain 
a ZGS at $3/4$ filling. Finally for pattern $A_7$ the distribution of ZGS is almost independent on both the anisotropy and dimerization.
The main message to retain from this Fig. \ref{zazbzc3}, is that the larger the dimerization the less important is the role of the anisotropy; 
in other words the strength of the dimerization along the different chains seems a more pertinent parameter than the anisotropy between the chains.

\subsubsection{k-space distribution of Dirac points associated to ZGS phases}

\begin{figure}

 \centering
\includegraphics[width=12cm]{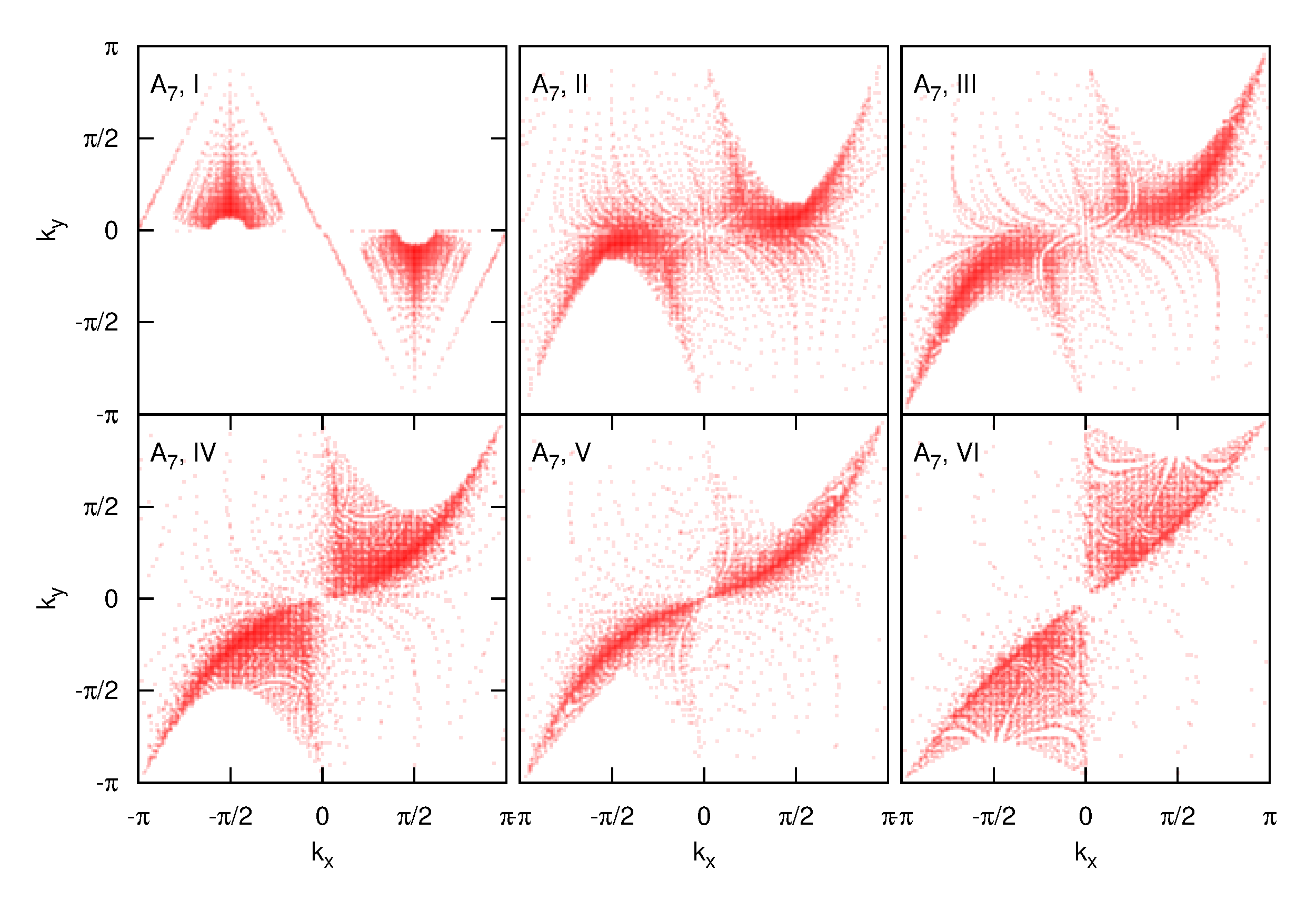}
 \caption{ k-space distribution of the Dirac points associated to the ZGS configurations obtained for patterns $A_7$ at $3/4$ filling
 for the six distinct dimerization configurations (I,II,...VI).}
\label{distriA7}
 \end{figure}

\begin{figure}

 \centering
\includegraphics[width=12cm]{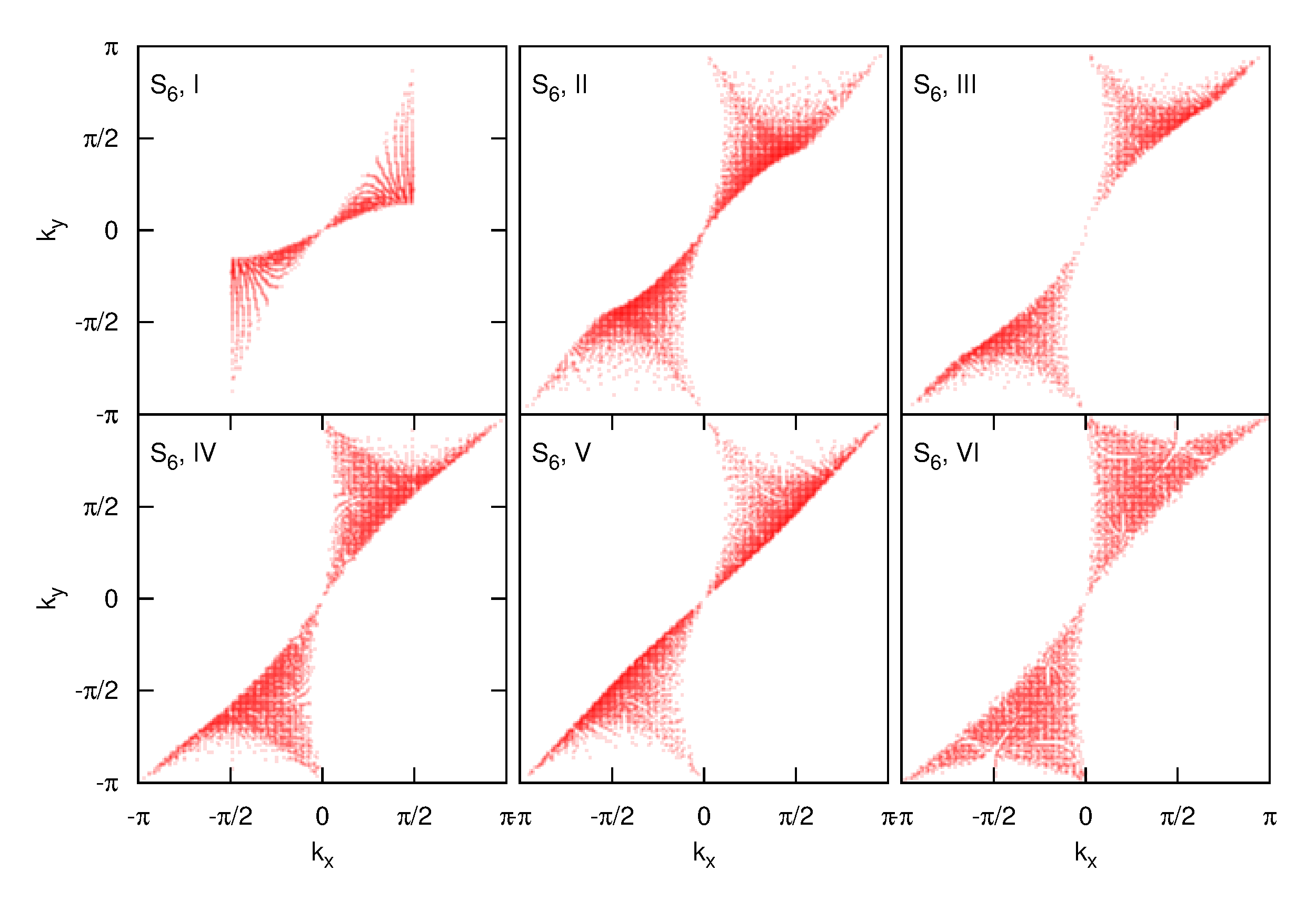}
\caption{ k-space distribution of the Dirac points associated to the ZGS configurations obtained for patterns $S_6$ at $3/4$ filling
 for the six distinct dimerization configurations (I,II,...VI).}
\label{distriS6}
 \end{figure}

As a last characterization of the ZGS properties, the following Fig. \ref{distriA3}, represents the evolution of 
the k-space distribution of the Dirac points associated to the ZGS at $3/4$ for pattern $A_3$ as function of the dimerization.
Figs. \ref{distriA7},\ref{distriS6} represent this same quantity but for pattern $A_7$ and $S_6$ respectively.
The most striking feature on these figures is that the case of no dimerization (I) is very different from all the other cases.
In fact, in the absence of dimerization the distribution of Dirac points associated to ZGS is much less spreaded over the Brillouin zone, moreover it differs strongly
from one pattern to the other. By increasing the dimerization, the k-space distribution of ZGS's Dirac points for patterns $A_3$ and $A_7$ become more and more similar, 
in fact they seem to only differ by number of Dirac points (intensity) as already seen in their corresponding histogram distribution. 
For sufficiently large dimerization (V,VI), the k-space distribution of ZGS's Dirac points for all patterns are very similar. All these features are further indications 
that the larger the dimerization the less important is the role played by the form of the chirality pattern.

\begin{figure}

 \centering
 \includegraphics[width=12cm]{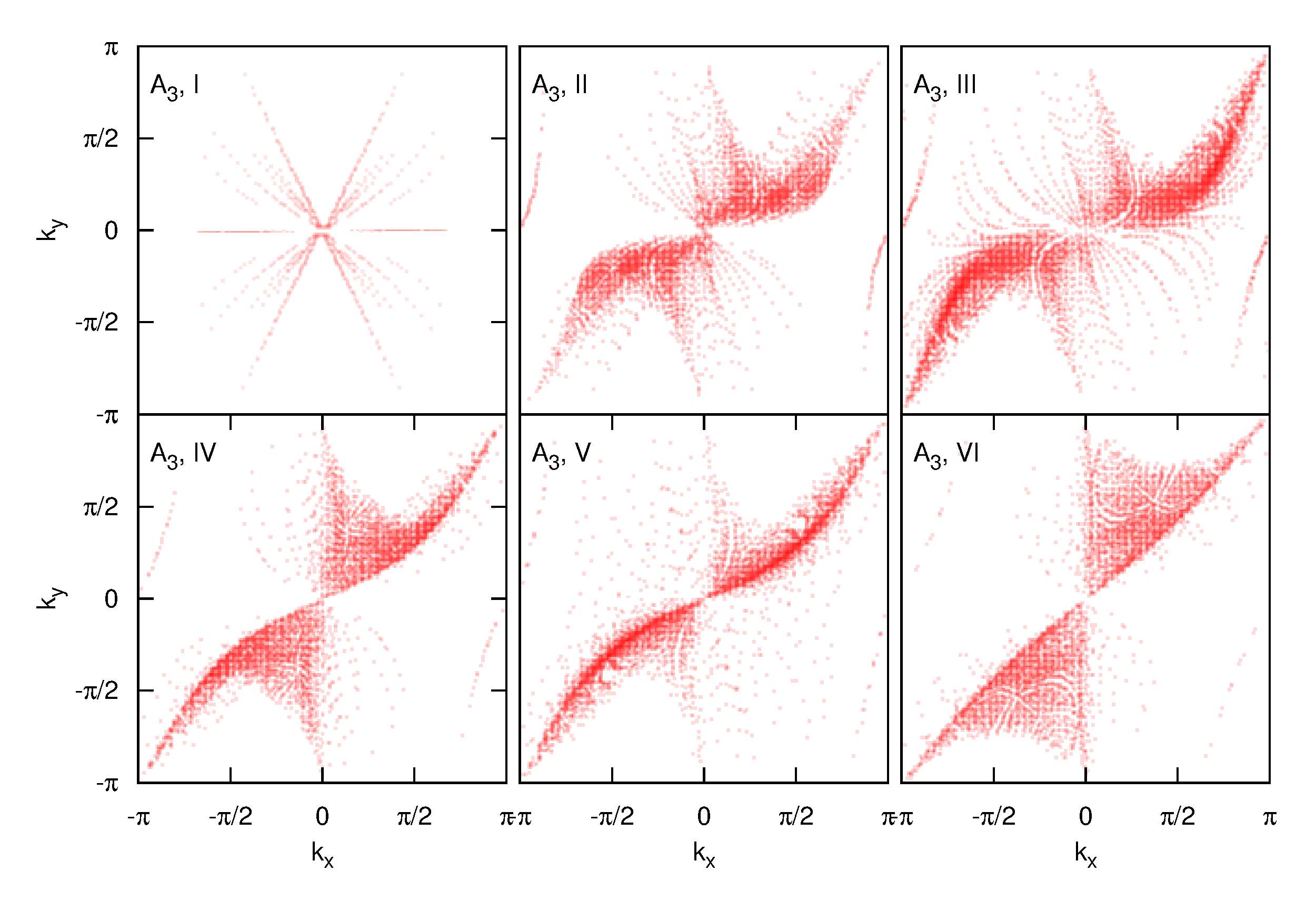}
 \caption{(color online) k-space distribution of the Dirac points associated to the ZGS configurations obtained for patterns $A_3$ at $3/4$ filling
 for the six distinct dimerization configurations (I,II,...VI).}
\label{distriA3}
 \end{figure}

\newpage

\section{Summary}

In summary, we have examined the existence of ZGS at $3/4$ filling for 
a tight binding model with seven nearest neighbors transfer integrals ($a_1, a_2, a_3, b_1, b_2, b_3, b_4$) between the four molecules of the unit cell, 
as in $\alpha$-type organic conductor.  We have systematically studied the combined role of the sign and moduli of the transfer integrals on the appearance of ZGS at $3/4$ filling.
We have shown that for given moduli, the are only 16 independent band structures $A_1, \cdots, S_8$. Each band structure is characterized 
by a chirality pattern $(\chi_1, \chi_2, \chi_3,\chi_4)$, where each chirality encodes the sign of the product 
of three transfer integrals that compose one of the four distinct triangular plaquette of the unit cell ($\chi_1 = {\rm sign} (a_2 b_1 b_4), 
\chi_2 = {\rm sign} (a_3 b_2 b_3), \chi_3 = {\rm sign} (a_1 b_1 b_2), \chi_4 = {\rm sign} (a_1 b_3 b_4)$).
In term of this classification, it appears that all previous works on $\alpha$-(BEDT-TTF)$_2$I$_3$ 
corresponds to pattern $A_3$ with $(\chi_1, \chi_2, \chi_3,\chi_4)\equiv(-,+,-,-)$.

As far as the properties of the LDOS and ZGS at $3/4$ (resp. $1/4$) filling are concerned, the 16 patterns can be further classified into only four different classes
determined by two effective chiralities $\chi_{a}={\rm sign} (a_1a_2)$ and $\chi_b={\rm sign} (b_1b_2b_3b_4)$.
Using extensive numerics we have established that for both large, or no inhomogeneity in the moduli $|a_1|,...|b_4|$, the
patterns $A_7,A_8$ (resp.  $A_1,A_2$) of class \suzu{$(\chi_a,\chi_b)=(-,-)$} are the most favorable patterns 
for the existence of ZGS at $3/4$ (resp. $1/4$) filling. At the opposite, for patterns ($S_1,S_2,S_7,S_8$) of class $(\chi_a, \chi_b)=$(+,+) 
we never found any ZGS at either $1/4$ or $3/4$ filling. Finally, patterns ($S_5,S_6$) (resp.  $S_3,S_4$) and $A_5,A_6$ (resp. $A_3,A_4$) of class
 \suzu{$(\chi_a, \chi_b)=(+,-)$} and \suzu{$(\chi_a, \chi_b)=(-,+)$} present an intermediate situation; they allow for ZGS at $3/4$ (resp. $1/4$) filling but are much less favorable
that patterns of class \suzu{$(\chi_a,\chi_b)=(-,-)$}. 
We note however that the role played by the chirality pattern is dependent on strength of the dimerization 
in the different effective chains formed by the transfer integrals. More precisely for weak or no dimerization the form of the chirality pattern strongly determines
the probability to find a ZGS at $1/4$ or $3/4$ filling whereas for strong dimerization 
 the possibility of having a ZGS become more or less equivalent for the three classes \suzu{$(-,-),(-,+)$ and $(+,-)$}. 
From that perspective we note that the usual transfer integral parameters taken to modelize $\alpha$-(BEDT-TTF)$_2$I$_3$ are in the strong dimerization regime.

On a larger perspective, we note that for bipartite lattice (like square lattice) when all plaquette chiralities are negative it corresponds to the so called uniform $\pi$-flux phase 
for which it is well known that there is a ZGS at $1/2$ filling. Our results suggest that on a non-bipartite lattice like the triangular lattice, 
it is necessary to have a fractional density $p/q$ of $\pi$-flux plaquettes to obtain a ZGS at some filling $r/q$; the difficulty in such a situation 
is that there appears many non equivalent chirality patterns.

\ack
\suzum{One of the authors (Y.S.) thanks G. Montambaux and  K. Awaga  for useful discussion.  
This work was supported 
 by a Grant-in-Aid for Scientific Research (A)
(No. 24244053) and 
(C)  (No. 26400355)
 from the Ministry of Education, Culture, Sports, Science, and Technology, Japan.
  }


\appendix
\section{Local density of states and chirality patterns}
\label{ldos}
Consider the $\alpha$-BEDT tight binding model with only nearest-neighbor hopping on the triangular lattice
\begin{equation}
H= 
\sum_{i}\sum_{<j,i>} \left[
  t_{ij} |i\rangle\langle j| +\rm{h.c}. \right],
 \label{eq:transfer_1} 
 \end{equation} 
 with $t_{i,j} \in (a_1,...,b_4)$.
The one particle local Greens function on a site $j$ is  $G_{jj}(E)=\langle j|(E-H)^{-1}|j\rangle$. It can be expanded as
\begin{equation}
 G_{jj}(E)=\frac{1}{E} \sum_{\ell} \frac{\langle j|H^{\ell}|j\rangle}{E^\ell}.
\end{equation}
The local density of states is obtained from the identity $D_{j}(E)=\frac{-1}{\pi} \Im m \lbrace G_{jj}(E+i0)\rbrace$,
which allows to write \suzum{\cite{LinNori1995,LinNori1996}}
\begin{equation}
 {\mathscr M}^\ell _j=\int_{- \infty} ^{\infty}{\rm{d}} E \ E^{\ell} D_{j}(E)=\langle j|H^{\ell}|j\rangle=
 \sum_{<i_1,j>}\sum_{<i_2,i_1>} ...\sum_{<i_{\ell-1},j>} t_{ji_1} t_{i_1i_2}...t_{\ell-1 j}.
\end{equation}
This last identity means that the $\ell^{th}$ moment ${\mathscr M}^\ell _j$ of the local density of states on site $j$ 
is equal to sum over all closed path of $\ell$ steps that start from $j$ and return to $j$. The key property is that
each path contributes with a weight proportionnal to the product of moduli of all the hoppings along the path and with a sign 
 given by the product of the chiralities of all the triangular plaquettes encircled by the path.
For the tight binding model of $\alpha$-BEDT, the moment for each site up to $\ell=3$ are easily obtained as:
\begin{equation}
\begin{array}{l}
  {\mathscr M}^2 _A=a_2 ^2+a_3^2+b_1^2+b_2 ^2+b_3^2+b_4^2,\\
  {\mathscr M}^2 _B=2(a_1 ^2+b_2 ^2+b_3^2),\\
  {\mathscr M}^2 _C=2(a_1 ^2+b_1 ^2+b_4^2),\\
  {\mathscr M}^3 _A=2(\chi_2|a_3b_2b_3|+\chi_1|a_2b_1b_4|)+\chi_3|a_1b_1b_2|+\chi_4|a_1b_3b_4|,\\
  {\mathscr M}^3 _B=2(\chi_2|a_3b_2b_3| +\chi_3|a_1b_1b_2|+\chi_4|a_1b_3b_4|),\\
  {\mathscr M}^3 _C=2(\chi_1|a_2b_1b_4| +\chi_3|a_1b_1b_2|+\chi_4|a_1b_3b_4|).\\
 \end{array}
\end{equation}
More generally, odd and even moment have the generic form:
\begin{equation}
\begin{array}{l}
 {\mathscr M}^{2\ell+1} _j=\sum_{\alpha} \chi_{\alpha} \ x^{\ell}_{\alpha,j}+
 \sum_{\alpha\ne \beta \ne \gamma} \chi_{\alpha}\chi_{\beta} \chi_{\gamma} \ x^{\ell}_{\alpha \beta \gamma,j},\\
  {\mathscr M}^{2\ell} _j=y^{\ell}_j+\sum_{\alpha\ne \beta} \chi_{\alpha}\chi_{\beta} \ y^{\ell}_{\alpha \beta,j}
  +\chi_1 \chi_2 \chi_3 \chi_4 \ y^{\ell}_{1234,j}
\end{array}
\end{equation}
where $(x^{\ell}_{\alpha,j},x^{\ell}_{\alpha \beta \gamma,j},y^{\ell}_j,y^{\ell}_{\alpha \beta,j},y^{\ell}_{1234,j})>0$ 
depend only on the moduli $(|a_1|,...,|b_4|)$.
For $|a_2|=|a_3|$, $|b_1|=|b_3|$ and $|b_2|=|b_4|$, one easily obtains that the global moments
${\mathscr M}^{\ell}=2{\mathscr M}^{\ell} _A+{\mathscr M}^{\ell} _B+{\mathscr M}^{\ell} _C$ verify
\begin{equation}
\begin{array}{l}
{\mathscr M}^{2\ell+1}=(\chi_1+\chi_2)(x^{\ell}_{1}+\chi_3\chi_4 x^{\ell}_{134})+(\chi_3+\chi_4)(x^{\ell}_{3}+\chi_1\chi_2 x^{\ell}_{123}),\\
{\mathscr M}^{2\ell}=y^{\ell}+(\chi_1+\chi_2)(\chi_3+\chi_4)y^{\ell}_{13}+\chi_1\chi_2 y^{\ell}_{12}+\chi_3\chi_4 y^{\ell}_{34}
+\chi_1 \chi_2 \chi_3 \chi_4 \ y^{\ell}_{1234}.
\end{array}
\label{A6}
\end{equation}

\section{The $8$ transfer integral configuration associated to pattern $A_3$}
\label{appendix_a}

\begin{table}
\caption{The $8=2^3$ configurations of signs of the seven transfer integrals that lead to the chirality pattern $A_3$.
All these configurations have the same LDOS $D_j(E)$. 
Strictly speaking for k-space dependent properties it is still necessary to distinguish two groups, i.e 
 configurations (1,2,5,6) with ($b_1b_4<0$ and $b_2b_3<0$) and congigurations  (3,4,7,8) with ($b_1b_4>0$ and $b_2b_3>0$).
 The energy bands $E_n({\bf k})$ of the first group  become equal to  that of the second group
 after the substitution $k_x \rightarrow k_x + \pi$. 
 }
\begin{center}
\begin{tabular}{cccccccccc}
\hline\noalign{\smallskip}
 $A_3$   & (1)& (2)& (3)& (4) & (5) & (6) & (7) & (8) \\
\noalign{\smallskip}\hline\noalign{\smallskip}
${\rm sign}(a _1)$  & + & + & + & + & - & - & - & - \\
${\rm sign}(a _2)$  & + & + & - & - & + & + & - & - \\
${\rm sign}(a _3)$  & - & - & + & + & - & - & + & + \\
${\rm sign}(b_1)$  &  + & - & + & - & + & - & + & - \\
${\rm sign}(b_2 )$  &  - & + & - & +& + & - & + & - \\
${\rm sign}(b_3 )$  &  + & - & - & + & - & + & + & - \\
${\rm sign}(b_4) $  &  - & + & + & - & - & + & + & - \\
\noalign{\smallskip}\hline
\end{tabular}
\end{center}
\label{table_a3}
\end{table}



\newpage


\medskip


\section*{References}

\begin{thebibliography}{9}




\bibitem{Mori1984_CL} 
\suzum{Mori T,  Kobayashi A,  Sasaki T, Kobayashi H,  Saito G, 
 and  Inokuchi H 1984 {\it Chem. Lett.}  \textbf{13}  957 
}

\bibitem{Katayama2006_JPSJ75} 
 Katayama S,  Kobayashi A, and  Suzumura Y 2006 
 {\it J. Phys. Soc. Jpn.} \textbf{75} 054705


\bibitem{Tajima2009_STAM10} 
 Tajima  N and Kajita K 2009 
 {\it Sci. Technol. Adv. Mater.} {\bf 10}  024308 

\bibitem{Kobayashi2009_STAM10}
 Kobayashi A,  Katayama S, and  Suzumura Y 2009 
{\it Sci. Technol. Adv. Mater.} {\bf 10} 024309

\bibitem{ZGS_review_2014} 
\suzum{Kajita K,  Nishio Y,  Tajima N,
  Suzumura Y, and  Kobayashi A 2014
 {\it J. Phys. Soc. Jpn.} \textbf{83} 072002}

\bibitem{Fu2007_PRB76} 
 Fu L and  Kane C L 2007 
 {\it Phys. Rev.} B {\bf 76} 045302

\bibitem{Piechon2013_JPSJ} 
 Pi\'echon F and  Suzumura Y 2013 
 {\it J. Phys. Soc. Jpn.} \textbf{82}  033703


\bibitem{Mori2013_JPSJ} 
 \suzum{Mori T  2013 {\it J. Phys. Soc. Jpn.} \textbf{82}  034714}


\bibitem{Mori2010_JPSJ} 
  Mori T  2010 {\it J. Phys. Soc. Jpn.} \textbf{79}  014701


\bibitem{Suzumura2013_JPSJ} 
 Suzumura Y,  Morinari T and  Pi\'echon F 2013 
 {\it J. Phys. Soc. Jpn.} \textbf{82} 023708

\bibitem{Piechon2014}
\suzum{Pi\'echon F,  Suzumura Y, and  Morinari T 2014 
 {\it presented at JPS September Meeeting, 2014}} 


\bibitem{Kondo2003}
 Kondo R,  Kagoshima S and  Maesato M 2003 
 {\it Phys. Rev.} B {\bf 67} 134519


\bibitem{Kondo2009}
 Kondo R,  Kagoshima S, Tajima N and  Kato R 2009
 {\it J. Phys. Soc. Jpn.} {\bf 78} 114714 

 
 
\bibitem{LinNori1995} 
 \suzum{ Nori F and  Lin Y-L 1994
 {\it Phys. Rev.} B {\bf 49} 4131 }


\bibitem{LinNori1996} 
 \suzum{Lin Y-L and  Nori F 1996  {\it Phys. Rev.} B {\bf 53} 13374
}



\end{thebibliography}
\end{document}